 \documentclass[final,5p,times]{elsarticle}

\usepackage{graphicx}
\usepackage{pstricks}
\graphicspath{{figures/}}

  \bibpunct{(}{)}{,}{a}{}{,}

\usepackage{epsfig}

\usepackage[switch,pagewise,running]{lineno}

\journal{Icarus}

\newcommand{\arcsec}{\ensuremath{^{\prime\prime}}}
\newcommand{\degr}{\ensuremath{^{\circ}}}

\newcommand{\ie}{\textsl{i.e.}}
\newcommand{\eg}{\textsl{e.g.}}

\usepackage[breaklinks]{hyperref}
\hypersetup{
    unicode=false,          
    pdftoolbar=true,        
    pdfmenubar=true,        
    pdffitwindow=false,     
    pdftitle={Disk-Resolved Spectroscopy of (1) Ceres},
    pdfauthor={Benoit Carry et al.},
    pdfsubject={Planetary Science},
    pdfkeywords={},         
    pdfnewwindow=true,      
    colorlinks=true,        
    linkcolor=gray,         
    citecolor=blue,         
    filecolor=gray,         
    urlcolor=gray           
}

\usepackage{ulem}
\newcommand{\add}[1]{\textbf{#1}}
\newcommand{\rem}[1]{\textcolor{red}{\sout{#1}}}

\renewcommand{\add}[1]{#1}
\renewcommand{\rem}[1]{}

\begin{document}

\begin{frontmatter}



\title{The remarkable surface homogeneity of the Dawn mission target (1) Ceres\tnoteref{eso-obs}}

\tnotetext[eso-obs]{%
  Based on observations collected at the European Organisation for
  Astronomical Research in the Southern Hemisphere, Chile - program
  ID: 
  \href{http://archive.eso.org/wdb/wdb/eso/eso_archive_main/query?prog_id=080.C-0881}{080.C-0881}.
}


\author[esac,lesia,eso-chile]{Beno\^{i}t Carry}
\author[eso-garching,lam,esa]{Pierre Vernazza}
\author[eso-chile]{Christophe Dumas}

\author[swri]{\add{William J. Merline}}

\author[besac]{Olivier Mousis}
\author[besac]{Philippe Rousselot}
\author[liege]{Emmanu\"{e}l Jehin}
\author[liege]{Jean Manfroid}
\author[lesia]{Marcello Fulchignoni}
\author[besac]{Jean-Marc Zucconi\fnref{dead}}

\address[esac]{European Space Astronomy Centre, ESA, P.O. Box 78, 28691 Villanueva de la Ca\~{n}ada, Madrid, Spain}
\address[lesia]{LESIA, Observatoire de Paris-Meudon CNRS, 5 place Jules Janssen, 92195 Meudon Cedex, France}
\address[eso-chile]{European Southern Observatory, Alonso de C\'{o}rdova 3107, Vitacura, Casilla 19001, Santiago de Chile, Chile}

\address[eso-garching]{European Southern Observatory, K. Schwarzschild-Str. 2, 85748 Garching, Germany}
\address[lam]{Laboratoire d’Astrophysique de Marseille, 38 rue Fr\'ed\'eric Joliot-Curie, 13388 Marseille, France}
\address[esa]{Research and Scientific Support Department, ESA, Keplerlaan 1, 2201 AZ Noordwijk, The Netherlands}

\address[swri]{Southwest Research Institute, 1050 Walnut St. \# 300, Boulder, CO  80302, USA}

\address[besac]{Universit\'{e} de Franche Comt\'{e} - CNRS, Institut UTINAM, 41 bis av. de l'Observatoire, F-25010 Besan\c{c}on Cedex, France}
\address[liege]{Institut d'Astrophysique et de G\'{e}ophysique,
  Universit\'{e} de Li\`{e}ge, B-4000 Li\`{e}ge, Belgium}

\fntext[dead]{Deceased}

\begin{abstract}

  \indent Dwarf-planet (1) Ceres is one of the two targets, along
  with (4) Vesta, that will be 
  studied by the NASA Dawn spacecraft via imaging, visible and
  near-infrared spectroscopy, and gamma-ray and neutron spectroscopy.
  \add{While Ceres' visible and near-infrared disk-integrated spectra have
    been well characterized, little has been done about quantifying 
    spectral variations over the surface. }
  Any spectral variation would give us insights
  on the geographical variation of the composition and/or the
  surface age. The only work so far was that of Rivkin \&
  Volquardsen (2010, Icarus 206, 327) who reported rotationally-resolved
  spectroscopic (disk-integrated) observations in the 2.2-4.0\,$\mu$m
  range; their observations showed evidence for a relatively uniform
  surface.\\
  \indent Here, we report disk-resolved observations of Ceres
  with SINFONI (ESO VLT) in the 1.17-1.32\,$\mu$m and 1.45-2.35\,$\mu$m
  wavelength ranges. The observations were made under
  excellent seing conditions (0.6\arcsec), allowing us to reach a
  spatial resolution of \add{$\sim$75\,km} on Ceres' surface. We do not find any
  spectral variation above a 3\% level, suggesting a 
  homogeneous surface at our spatial resolution.
  \add{Slight variations (about 2\%)
    of the spectral slope are detected, geographically
    correlated with the albedo
    markings reported from the analysis of the HST and Keck
    disk-resolved images of Ceres
    (Li et al., 2006, Icarus 182, 143; Carry et al., 2008, A\&A 478,
    235).
    Given the lack of constraints on the surface composition of Ceres,
    however, we cannot assert the causes of these variations.}
\end{abstract}


\end{frontmatter}

   \pagewiselinenumbers


\section{Introduction}

  \indent Ceres, discovered in 1801 by Giuseppe Piazzi, is
  - with a diameter of 935\,km \citep{2008-AA-478-Carry} -
  by far the largest asteroid in the main
  asteroid belt. Interestingly, recent simulations
  \citep[\eg,][]{2009-Icarus-204-Morbidelli, 2009-Nature-457-Minton} 
  have shown that in the early Solar System there must have been
  many Ceres-sized bodies
  \rem{, up to a few thousand depending on assumptions,}
  to explain the current size-frequency distribution of
  asteroids.
  Ceres therefore appears to be a unique
  remnant of a dynamical removal process, which caused the 
  Main Belt to lose \add{most} \rem{99.9\%} of its primordial mass.
  By furthering our understanding of the physical properties
  (\eg, surface composition, \add{density, internal structure})
  of the largest asteroids, we may be afforded a
  special opportunity
  to learn more about the
  formation process of large planetesimals from which terrestrial
  planets once accreted.\\
  \indent As one of the targets of the NASA Dawn mission
  \citep[see][for instance]{2004-PSS-52-Russell}, Ceres
  recently has been the subject of increased attention from the
  community
  \citep[\eg,][and references therein]{
    2002-AJ-123-Parker,
    2003-AdSpR-31-Nazzario,
    2005-Nature-437-Thomas,
    2005-AA-436-Vernazza,
    2005-JGR-110-McCord, 
    2006-Icarus-182-Li,
    2006-Icarus-185-Rivkin,
    2007-Icarus-188-Chamberlain,
    2007-EMP-100-Kovacevic,
    2008-Icarus-197-Drummond,
    2008-MNRAS-383-Mousis,
    2009-NatGe-2-Milliken,
    2009-Icarus-202-Chamberlain,
    2010-Icarus-205-Castillo-Rogez,
    2010-AA-516-Moullet,
    2011-AJ--Rousselot}.
  Three topics seem to have drawn the most attention:
  1) characterization of the surface composition,
  2) thermodynamic modeling of the interior,
  and
  3) determination of the
  size, shape, spin, and albedo
  via disk-resolved imaging.
  We summarize the main results here:

  \begin{enumerate}
    \item While Ceres' exact surface composition remains elusive
      \citep[see][for a review]{2010-SSRv--Rivkin}, the
    analysis of the 2.9-4.0\,$\mu$m region has allowed the possible
    detection of brucite and carbonate assemblages on its surface
    \citep{2009-NatGe-2-Milliken}, suggesting
    that the thermal and aqueous alteration history of
    Ceres differs from that in the record of carbonaceous meteorites.
    
    \item Modeling of Ceres' thermo-physical-chemical evolution
    has pointed to a 
    differentiated interior for Ceres, 
    in accord with shape measurements
    \citep{2005-Nature-437-Thomas, 2008-AA-478-Carry}.
    The interior would consist of
    an inner core of dry silicates, an intermediate layer of
    hydrated silicates, and an outer shell of water ice
    \citep{2010-Icarus-205-Castillo-Rogez}.
    However, \add{an} undifferentiated Ceres consisting of a
    homogeneous mixture of hydrated silicates is still possible
    \citep{2009-Icarus-204-Zolotov}.
    
    \item Disk-resolved imaging of Ceres has allowed determination of
    its size, shape, and spin with small uncertainties, in turn
    constraining the density
    \citep{2005-Nature-437-Thomas, 2008-AA-478-Carry}.
    Finally, the images showed surface albedo variations of 6 \% around the
    average surface value, with putative, broad-band, color variations
    \citep[$\approx$2--3\%, see][]{2006-Icarus-182-Li, 2008-AA-478-Carry}.
    The cause of the
    color contrast (\eg, geological features, composition, grain
    properties, surface age) could not be determined. 
 
  \end{enumerate}

  \indent While Ceres' visible and
  near-infrared spectrum has been well characterized, little has
  been done about quantifying 
  spectral variations over the surface.
  Disk-resolved analysis would give us insights on the
  geographical variation of the composition and/or variation in
  ages of surface units.
  The only disk-resolved analyses done so far were those of
  \citet{2002-AJ-123-Parker}, 
  \citet{2006-Icarus-182-Li}, and
  \citet{2008-AA-478-Carry}, studying the albedo/color
  variations in the visible and near-infrared only.
  Observations at other
  wavelegths were limited to disk-integrated observations and could only
  search for rotational modulation of the spectra.
  All studies showed evidence for a relatively uniform
  surface.
  A summary of the disk-integrated results is given here:

\begin{enumerate}
  \item Polarization:
    no rotational variation reported \add{at visible wavelengths}
    \citep{2010-SSRv--Rivkin}, although a \add{slight}
    variation of 0.1\% in the
    degree of polarization can be surmised from the literature
    (I. Belskaya, personnal communication)
  \item Visible photometry: low-amplitude lightcurve due to albedo
    features \citep{1983-Icarus-54-Tedesco}.
  \item Near-infrared spectroscopy:
    marginal (1\,$\sigma$) variations of brucite- and
    carbonates-associated
    absorption bands \citep{2010-Icarus-206-Rivkin}, indicating
    a homogeneous mineralogy with tenuous evidence for variations.
  \item Far infrared radiometry: very low-amplitude lightcurve
    \citep{1996-AA-309-Altenhoff,2010-AA-516-Moullet}, indicating
    an overall homogeneous surface and subsurface properties
    (roughness, absorption coefficient, refractive index, and thermal inertia).
\end{enumerate}

  \indent Although the disk-integrated spectrum of Ceres presents more 
  absorption features in the 2.9-4.0\,$\mu$m region
  \add{\citep[\eg,][and references therein]{2010-SSRv--Rivkin}},
  we report an analysis of ground-based disk-resolved
  spectral observations of Ceres' surface in the near-infrared
  (\add{1.1}-2.4\,$\mu$m) to
  investigate with better spectral resolution the surface features
  detected by \citet{2008-AA-478-Carry} from disk-resolved, broad-band,
  imaging observations.
  At this time, high-angular-resolution imaging
  spectrographs can only operate in the near-infrared range.
  In Section~\ref{sec: obs} we describe the observations and the data
  reduction,
  in Section~\ref{sec: results} we present
  the spectral analysis, and
  in Section~\ref{sec: discussion} we
  discuss the spectral heterogeneity across Ceres' surface.

\section{Observations and Data Reduction\label{sec: obs}}

  \indent We made disk-resolved observations of (1) Ceres on
  2007 November 13, covering about 2/3 of its 9.074 h
  rotation \citep{2007-Icarus-188-Chamberlain, 2008-AA-478-Carry}.
  The observations were made in the
  near-infrared with the integral-field spectrograph
  SINFONI \add{\citep[using the J: 1.17-1.32\,$\mu$m and H+K:
      1.45-2.35\,$\mu$m gratings, with a resolving power of 2000 and
      1500 respectively, see][]{2003-SPIE-1548-Eisenhauer,
      2004-Msngr-117-Bonnet}}
  at the European Southern Observatory (Cerro Paranal, Chile) with one
  (UT4/Yepun) of the four 8-m telescopes of the Very Large Telescope
  (VLT) facility. \\ 
  \indent \add{%
    Atmospheric conditions were excellent, with
    a seeing at visible wavelengths of about 0.6\arcsec\,(reaching
    1.1\arcsec\,at worst) and a coherence time of 2--4\,ms.
    The adaptive optics module of SINFONI was providing a
    high-quality correction:
    the wavefront sensor used Ceres itself (V\,$\sim$\,7.3) as a
    reference, allowing high-frequency correction (420\,Hz).
    Identical settings were used for the observation of the
    standard stars (see below), and the quality of the correction was
    similar in both cases.} \\
  \indent The apparent diameter of Ceres at the time of the observation
  was about 0.7\arcsec, so to search for spectral variations across
  the asteroid surface, we 
  used the high-angular-resolution mode with an equivalent pixel
  size on sky of 25$\times$12.5 milli-arcsec and a field of view of
  0.8$\times$0.8 arcsecond.
  The high angular resolution, resulting from the adaptive-optics
  correction coupled with the excellent seeing conditions we had
  throughout the night (see Table~\ref{tab: obs}),
  allowed us to reach an average spatial resolution on Ceres' surface of
  \add{about 75\,km (ranging from 55 to 98\,km, depending on the
  wavelength, see Table~\ref{tab: psf}).}
  The same jitter pattern was used at all epochs to ensure complete
  coverage of Ceres' apparent disk given the narrow field of view:
  a single exposure of Ceres is made at the center of
  the pattern, followed by four others, taken in (x,y) steps of $\pm$ 0.2 arcsec
  from the central position. Then
  three sky exposures are taken 30\arcsec~away from Ceres (East - 
  North - West).
  Exposure times were 90\,s and 40\,s in J and H+K, respectively
  (we used 4\,s and 1\,s, respectively, for the standard stars). \\
%
\begin{table}[!ht]
\begin{center}
\begin{tabular}{ccrrccc}
\hline
\hline
Time & Grating & \multicolumn{1}{c}{SEP$_\lambda$} &\multicolumn{1}{c}{SSP$_\lambda$} & X & $s$ \\
(UT) &         & \multicolumn{1}{c}{(\degr)}      & \multicolumn{1}{c}{\degr)} & & (\arcsec) \\
\hline
00:48 & H+K & 211 & 213 & 2.04 &  0.79 \\
00:58 & H+K & 204 & 206 & 1.92 &  0.67 \\
01:12 &  J  & 195 & 197 & 1.77 &  1.28 \\
01:55 & H+K & 167 & 169 & 1.49 &  0.73 \\
02:05 & H+K & 160 & 162 & 1.44 &  1.10 \\
02:20 &  J  & 150 & 152 & 1.37 &  1.09 \\
02:57 & H+K & 125 & 127 & 1.27 &  0.61 \\
03:07 & H+K & 119 & 121 & 1.25 &  0.57 \\
03:18 &  J  & 111 & 113 & 1.23 &  0.61 \\
03:57 & H+K &  86 &  88 & 1.19 &  1.10 \\
04:07 & H+K &  79 &  81 & 1.19 &  1.11 \\
04:18 &  J  &  72 &  74 & 1.19 &  1.10 \\
04:37 & H+K &  60 &  62 & 1.19 &  1.04 \\
04:46 & H+K &  53 &  55 & 1.20 &  n.a. \\
04:58 &  J  &  46 &  47 & 1.21 &  0.86 \\
05:16 & H+K &  34 &  36 & 1.23 &  0.72 \\
05:26 & H+K &  27 &  29 & 1.25 &  0.78 \\
05:37 &  J  &  20 &  22 & 1.27 &  0.62 \\
06:15 & H+K & 355 & 356 & 1.38 &  0.75 \\
06:25 & H+K & 348 & 350 & 1.42 &  0.60 \\
06:36 &  J  & 341 & 343 & 1.47 &  0.56 \\
07:10 & H+K & 318 & 320 & 1.68 &  0.45 \\
\hline
\end{tabular}
\caption[Observing circumstances]{%
  Observing circumstances for (1) Ceres.
  For each observation, we report the
  mid-observation time in UT,
  the grating used to disperse the light,
  the \add{longitude ($\lambda$)} of the 
  sub-Earth point (SEP) and subsolar point (SSP),
  \add{their respective latitude $\beta$ being
    3\degr~and 0\degr,}
  the airmass (X), and the seeing ($s$).
\label{tab: obs}
}
\end{center}
\end{table}
%
  \indent The observing circumstances are reported in Table~\ref{tab: obs}.
  \add{We list} the mid-observation time (UT),
  grating used (J or H+K), 
  longitude of the sub-Earth and subsolar points
  (SEP$_\lambda$ and SSP$_\lambda$, respectively,
  \add{computed using the pole solution by \citet{2008-AA-478-Carry}}),
  airmass (X), \add{and seeing ($s$).}
  During our observations, the latitude of the SEP and
  SSP was constant, of 3\degr~and 0\degr,~respectively. \\
  \indent Observations of standard (PSF) stars were interspersed with Ceres
  observations throughout the night. 
  Two stars were used:
  HD~15474 (G5) and HD~28099 (G2V).
  \add{Both stars have V\,$\sim$\,8.1, slightly dimmer than
    Ceres (V\,$\sim$\,7.3 at the time of the observations), and were
    located at
    8.5\degr~and
    22\degr~from Ceres respectively.}
  We list in
  Table~\ref{tab: psf} the circumstances for those observations, namely
  the mid-observation time (UT), 
  the grating used (J or H+K), the airmass (X),
  and
  the angular resolution measured on the star itself after data
  reduction ($\Theta$, corresponding to its full width at half maximum, FWHM).
  \add{Although the seeing and airmass varied thorough the night, the
    angular resolution provided by the AO correction was very stable
    (Table~\ref{tab: psf}). We therefore consider that the spatial
    resolution at Ceres was 55, 71, and 98\,km in the J, H, and K-band
    wavelength ranges (the resolution decreasing continuously from the
    shortest wavelength to the longest).}

%
%
%
%
%
%
 
%
%
%
\begin{table}
\begin{center}
\begin{tabular}{lcccccc}
\hline
\hline
\multicolumn{1}{c}{Time} & Grating & X & $s$ & $\Theta_J$ & $\Theta_H$ & $\Theta_K$ \\
\multicolumn{1}{c}{(UT)} & & &(\arcsec) &  (\arcsec) &(\arcsec) & (\arcsec)  \\
\hline
00:34          &  J  & 1.82 & 0.72 & 0.030 &  --   &  --   \\
01:44          &  J  & 1.39 & 1.33 & 0.042 &  --   &  --   \\
02:47          &  J  & 1.23 & 0.75 & 0.045 &  --   &  --   \\
03:47          &  J  & 1.19 & n.a. & 0.042 &  --   &  --   \\
06:05$^\dagger$ &  J  & 1.34 & 0.58 & 0.047 &  --   &  --   \\
07:02          &  J  & 1.92 & 0.53 & 0.043 &  --   &  --   \\
\noalign{\smallskip}
00:23          & H+K & 1.92 & 0.64 &  --   & 0.049 & 0.076 \\
01:36          & H+K & 1.42 & 1.24 &  --   & 0.054 & 0.069 \\
02:41          & H+K & 1.24 & 0.73 &  --   & 0.052 & 0.070 \\
03:41          & H+K & 1.19 & n.a. &  --   & 0.057 & 0.068 \\
05:58$^\dagger$ & H+K & 1.34 & 0.54 &  --   & 0.054 & 0.080 \\
06:56          & H+K & 1.87 & 0.55 &  --   & 0.053 & 0.076 \\
\noalign{\smallskip}
&&\multicolumn{2}{c}{Average}     & 0.041 & 0.052 & 0.073 \\
&&\multicolumn{2}{c}{Theoretical} & 0.031 & 0.041 & 0.057 \\
&&\multicolumn{2}{c}{Deviation}   & 0.006 & 0.003 & 0.005 \\
\hline
\end{tabular}
\caption[Observing circumstances for standard stars.]{%
  Observing circumstances for standard stars.
  For each observation, we report the
  mid-observation time in UT,
  the airmass (X), the seeing ($s$), 
  \add{and the size of the resolution element, measured as the FWHM of
    the stars ($\Theta$), for each band: J, H, and K (\ie, at 1.22, 1.60, and
  2.2\,$\mu$m).
  The theoretical angular resolution (defined here as the diffraction
  limit $\lambda$/$D$, with $\lambda$ the wavelength and $D$ the
  telescope aperture) is also reported.
  The average FWHMs of the PSFs correspond to spatial resolutions of
  55, 71, and 98\,km respectively (projected to the distance of Ceres).}
  The observation marked with a dagger ($^\dagger$) corresponds to the
  single observation of HD~28099. 
\label{tab: psf}
}
\end{center}
\end{table}

\begin{figure}[!ht]
  \includegraphics[width=.4\textwidth]{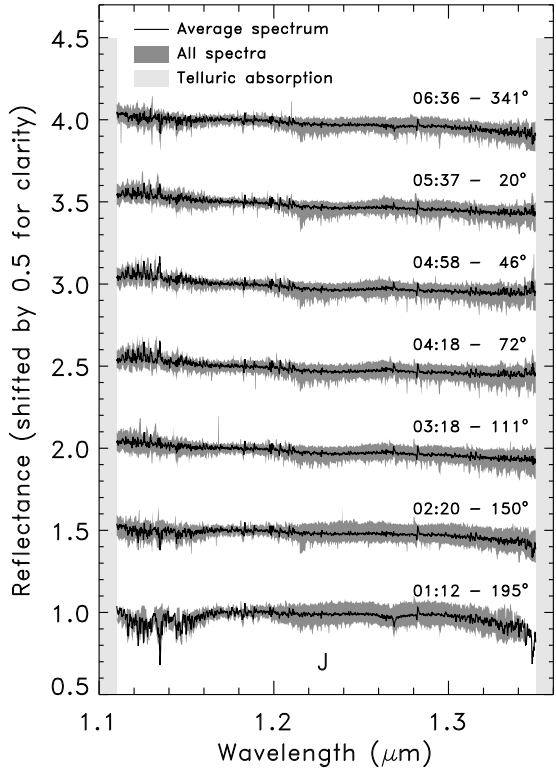}
  \caption[Spectra obtained with the J grating]{%
    Reflectance spectra obtained with the J grating.
    For each epoch, the average spectrum (black line)
    and all the individual spectra within the ROI (dark-gray area)
    are plotted.
    The wavelength ranges affected by telluric absorptions
    (\ie, the limits of the J-band)
    are displayed in light gray
    \add{and the spectra are omitted for clarity}.
    The mid-observation time \add{(UT time on 2007 Nov 13)} and
    the longitude of the sub-Earth point
    (SEP$_\lambda$, see Table~\ref{tab: obs}) are reported for each epoch.
\label{fig: J}
}
\end{figure}

\begin{figure}[!ht]
  \includegraphics[width=.4\textwidth]{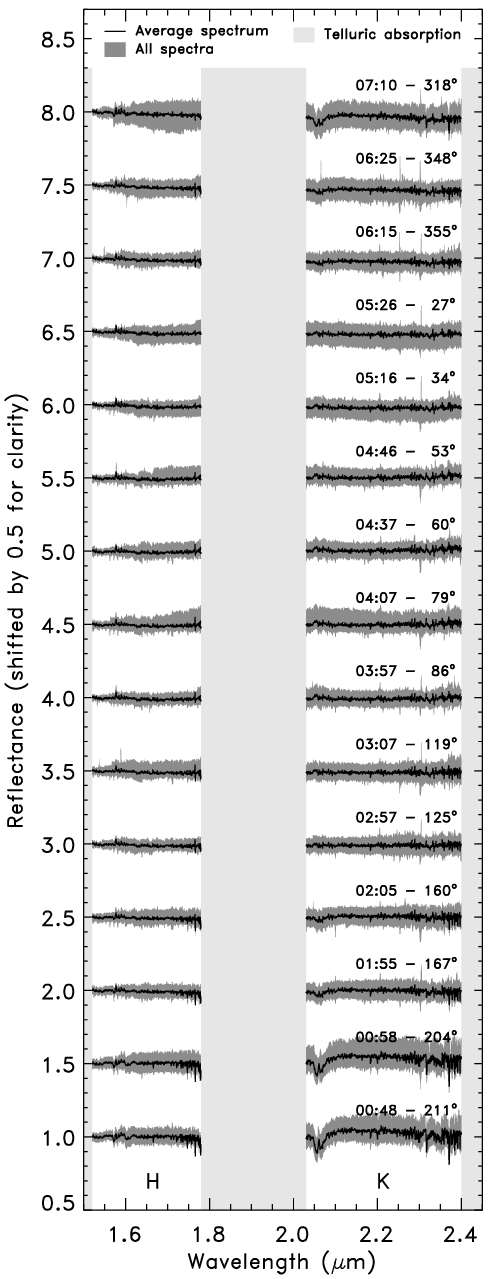}
  \caption[Spectra obtained with the H+K grating]{%
    Reflectance spectra obtained with the H+K grating.
    For each epoch, the average spectrum (black line)
    and
    all the individual spectra with the ROI (dark-gray area)
    are plotted.
    The wavelength ranges affected by telluric absorptions
    (\ie, the limits of the H- and K-bands)
    are displayed as the light gray area
    \add{and the spectra are omitted for clarity}.
    The mid-observation time  \add{(UT time on 2007 Nov 13)}
    and the longitude of the sub-Earth point
    (SEP$_\lambda$, see Table~\ref{tab: obs}) are reported for each
    epoch.
    Absorption features at 2.05\,$\mu$m visible in spectra taken at
    00:48, 00:58, and 07:10 are telluric absorption not fully-corrected due to
    the high airmass of Ceres during the observations.
\label{fig: HK}
}
\end{figure}

  \indent The basic data reduction was performed using the ESO 
  pipeline 2.0.5 \citep{2007-arXiv-Modigliani}:
  1) correction of the object-sky pairs for bad pixels and
  flat-field effects,
  2) wavelength calibration,
  3) sky subtraction, and
  4) reconstruction of the final image cubes (the 3-dimensional
  ``cubes'' consist of many 2-dimensional images of Ceres; the third
  dimension is wavelength, with one image for each wavelength channel).
  As already discussed by
  \citet{2010-Icarus-205-Carry-b}, the grazing viewing
  angle near Ceres' limb causes unreliable
  flux measurement for the outer annulus of the asteroid's apparent
  disk.
  Therefore, we restricted our study to a ``region of interest''
  (ROI), covering the innermost portion of the apparent disk. The ROI was
  defined to be the central region of the disk, within 75\% of Ceres' radius,
  \citep[following][]{2008-AA-478-Carry}.\\
  \indent Each spectrum extracted within the ROI was divided by
  the spectrum of a standard star, taken close in time and in airmass
  (see Table~\ref{tab: obs} and Table~\ref{tab: psf}), to ensure the
  best possible correction for atmospheric effects, particularly water vapor.
  \add{Remaining bad pixels and cosmic rays were removed by applying a
  median filter to the reflectance spectra.
  The box size was set to 5 pixels,
  corresponding to 0.7 and 2.5\,nm for J and H+K gratings respectively.
  This width is narrow enough to have no influence on the detection of
  absorption bands (resolving power remains of several hundreds.}
  \rem{These reflectance spectra were smoothed with a median filter,
  using a box size of 5 pixels,
  corresponding to 0.7 and 2.5 nm for J and H+K gratings, respectively. }
  The spectra are
  normalized to unity at 1.17\,$\mu$m and \add{1.50}\,$\mu$m, respectively.
  These reflectance spectra are plotted for each epoch in
  Figs~\ref{fig: J} and~\ref{fig: HK}.

\begin{figure}[!ht]
  \includegraphics[width=.49\textwidth]{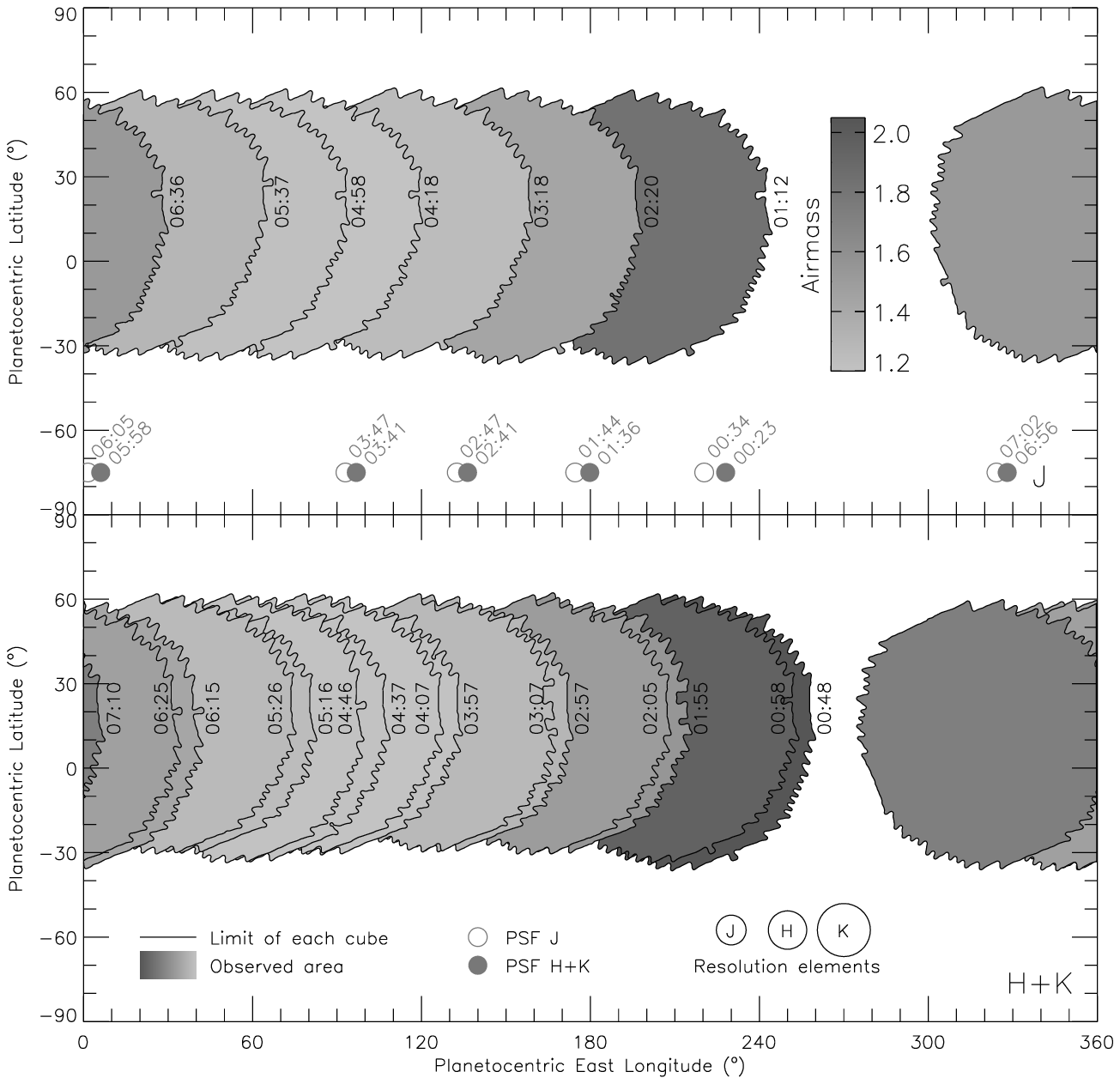}
  \caption[Map showing the coverage of the observations]{%
    Equidistant cylindrical projections showing the coverage of our
    observations (ROI only) on Ceres' surface.
    Each image cube is delimited by a black contour \add{and
      shaded as a function of airmass at the time of the
      observation (Table~\ref{tab: obs})}.
    The observing time of each image cube is reported near its eastern border.
    \textsl{Top}: the 7 image cubes using the J grating. 
    Filled and empty circles (arbitrary size),
    labeled with UT times, \add{indicate} the
    rotational phase of Ceres at the observation time of the standard
    stars with the H+K and J grating, respectively.
    \textsl{Bottom}: the 15 image cubes using the H+K grating. 
    \add{The size of the resolution elements at each wavelength range
      (55, 71, and 98\,km at J, H, and K-band) are represented here}, as
    if projected on Ceres equator.
\label{fig: cov}
}
\end{figure}

\section{Spectral analysis\label{sec: results}}

  \indent Our observations covered most of Ceres's low-latitude regions
  (see Table~\ref{tab: obs}, and Fig.~\ref{fig: cov}).  
  In both wavelength ranges, we observe very little deviation from
  the average (see Figs~\ref{fig: J} and~\ref{fig: HK}).
  To quantify precisely how the different reflectance spectra
  behave with respect to the average, we utilized the two following
  norms, sensitive
  to spectral deviation (\ie, presence of absorption band),  
  that we applied to all the spectra of each \add{image-}cube
  (both J and H+K; spectra within the ROI)
  before (case $a$) and after (case $b$) division by their spectral
  slope ($\approx$continuum\add{, re-normalized to unity at
    1.17\,$\mu$m and 1.50\,$\mu$m, see Sect.~\ref{sec: obs}}): 
  
\begin{equation}
  \sigma_i = \frac{ \sum_\lambda\left( \mathcal{S}_i-1 \right)}
                { \sum_\lambda \left( \mathcal{S}^{\star}-1 \right) }
  \label{eq: dev}
\end{equation}

\begin{equation}
  \chi_i^2 = \frac{\sum_\lambda (\mathcal{S}_i-1)^2}
      {\sum_\lambda (\mathcal{S}^{\star}-1)^2}
  \label{eq: chi}
\end{equation}

  \noindent where 
  $\sigma_i$ and $\chi_i^2$ are the reduced ``\textsl{deviation}'' and
  ``\textsl{chi-square}'' norms,
  $\mathcal{S}^\star$ the slope-removed spectrum used to estimate the
  noise,  
  $\mathcal{S}_i$ the spectrum normalized by the disk-integrated
  average spectrum (all the spectra within the ROI);
  the subscript $i$ standing for cases $a$ and $b$.\\
  \indent Because both norms are reduced
  (\ie, they are normalized by a measure of the noise), 
  they give the level of confidence at which a variation from the
  disk-integrated average spectrum is detected:
  a spectrum equal to the average spectrum at all wavelengths has a
  norm of 0, and 
  a spectrum whose deviations to the average spectrum are smaller than
  its intrinsic noise will have a norm below 1.

  We applied the norms before ($a$) and after ($b$)
  slope removal to test if the
  spectral variation was mainly a slope variation or if it could
  also be due to the presence of absorption bands.

\begin{figure}[!ht]
  \includegraphics[width=.49\textwidth]{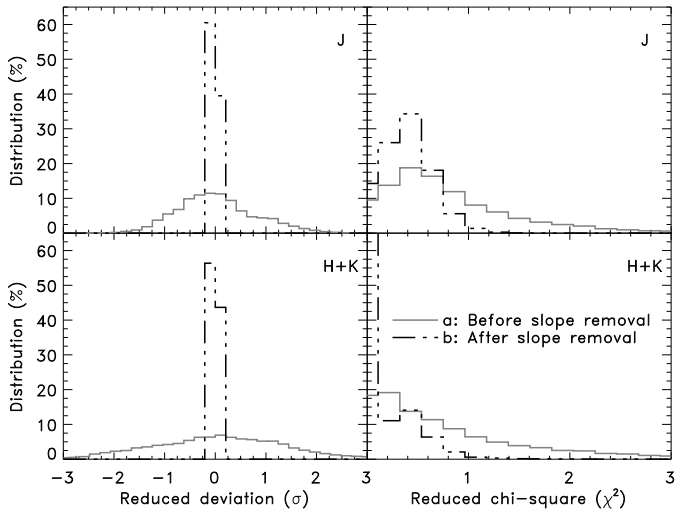}
  \caption[Variation distributions]{%
    Reduced deviation ($\sigma$, left) and
    chi-square ($\chi^2$, right) norms distribution
    for both J (top) and H+K (bottom) spectra, 
    before (case $a$, gray histogram) and 
    after (case $b$, black histogram) slope removal.
\label{fig: histo}
}
\end{figure}

\begin{figure*}[!ht]
\begin{center}
  \includegraphics[width=.75\textwidth]{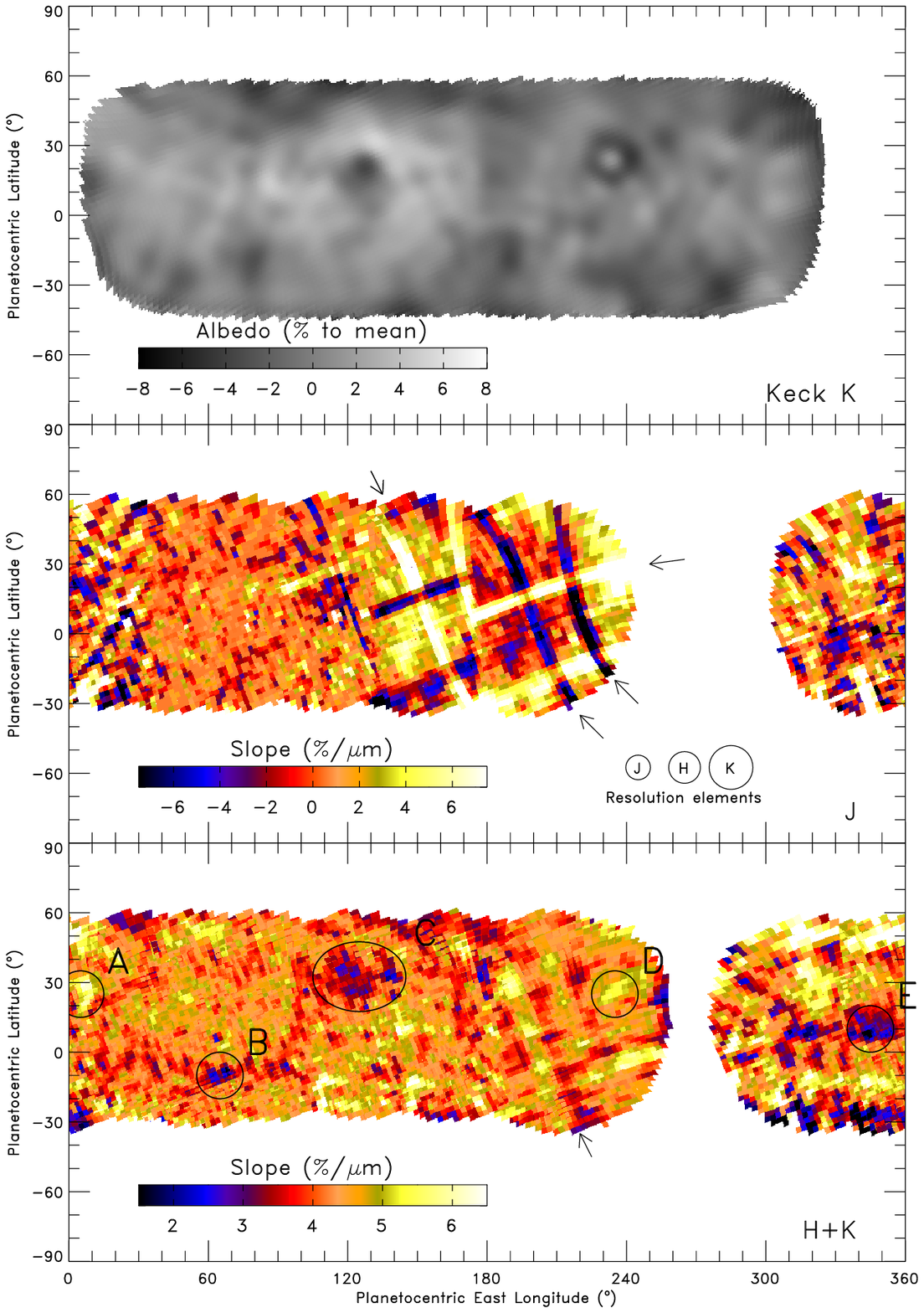}%
  \caption[Map of the spectral slope distribution]{%
    Equidistant Cylindrical Projection maps of the spectral slope
    measured in J (middle) and H+K (bottom) wavelength ranges,
    compared with the K-band albedo map (top) from 
    \citet{2008-AA-478-Carry}.
    \rem{The distribution of variations is reproduced from cube to cube and
    is therefore associated to noise patterns rather than surface
    variations. }
    \add{The size of the resolution elements at each wavelength range
      (55, 71, and 98\,km at J, H, and K-band) are represented here
      (middle map), as if projected on Ceres' equator.
      No features above the noise level are reliably seen in the J-band
      map, and most of the small-scale features seen in H+K are likely
      noise.
      There are five regions where the slope deviates from the
      average (about 4\%/$\mu$m in H+K),
      labeled from A to E on the H+K map.
      The main uncorrected noise patterns (stripes crossing the field of
      view) are highlighted with arrows (see text).} 
\label{fig: mapslope}}
\end{center}
\end{figure*}

  \indent We display in Fig.~\ref{fig: histo}
  the distribution of these two norms, computed
  for the 10\,619 and 22\,755 individual spectra obtained with
  the J and H+K gratings respectively.
  First, both distributions are
  sharper, and mostly below 1,
  after slope removal (case $b$)
  than before (case $a$).
  This computation highlights the \add{small} spectral variation,
  mostly due to slope variation,
  of each spectrum with respect to the average one.\\
  \indent We then focused our attention on spectra with $\chi_b^2$ norm above
  1, to check whether the variation is real or due to an instrumental
  artifact, and if real whether the variation is due to the repeated
  presence of an absorption band. 
  Reflectance spectra with higher norms appeared at fixed positions on
  the SINFONI
  detector, instead of following the rotation of Ceres. We thus
  conclude on the absence of absorption band in J, H and K-band
  wavelengths at our noise level (1-$\sigma$):
  3.7\% in J, 1.4\% in H, and 1.8\% in K. 
  \rem{This translates into an upper limit of approximately 1\% for ice
  frost geometric coverage.}
  \add{Considering the depth of the characteristics 
    absorption bands of water ice, centered at 1.6 and 2.2\,$\mu$m,
    this sets an upper limit of
    approximately 1\% for ice frost geometric coverage.}\\
  \indent We finally investigated the spatial distribution of the
  spectral slope variation. 
  For each spatial pixel, we computed the spectral slope, and \add{mapped}
  it onto an Equidistant Cylindrical Projection (ECP) 
  (Fig~\ref{fig: mapslope}),
  following \citet{2010-Icarus-205-Carry-b}.
  \add{The slope variations were reproduced from image cube to
    image cube, and were thus linked to the instrument.
  We corrected this cosmetic \add{problem}
  by subtracting an average slope map at each image cube.
  The first observations of the night were more affected by this
  noise pattern, and some residuals are still present in 
  the slope maps presented in Fig.~\ref{fig: mapslope},
  between 120\degr~and 240\degr~East in J,
  and between 180\degr~and 250\degr~East in H+K
  (corresponding to high-airmass observations,
  see Fig.~\ref{fig: cov}
  and Table~\ref{tab: obs}).
  The slope information at these location is not reliable, and
  interpretations must be cautious. } \\
  \indent \add{No slope variations are detected in the J-band
    wavelength range, all the deviations to the average spectral slope
    (neutral between 1.10 and 1.35\,$\mu$m) appearing suspicious.
    We, however, detect possible genuine variations (slope only) over
    the H+K wavelength range in
    five regions, labeled A to E (Fig.~\ref{fig: mapslope}).
    Three of them (B, C, and E) present a lower spectral slope (about
    2\%/$\mu$m) than the surroundings
    (the average spectral slope of Ceres in that wavelength range is
    of about  4\%/$\mu$m).
    These location are associated with low-albedo markings at
    visible
    and near-infrared wavelengths:
    feature \#7, central part of \#2, and East of \#6 in Fig.~5
    by \citet{2006-Icarus-182-Li}
    and 
    features d$_1$ and d$_3$ in Fig.~7 by \citet{2008-AA-478-Carry}.
    The other two features (A and D) have higher spectral slopes
    (above 5\%/$\mu$m) and are associated with high albedo markings:
    features \#1 and \#5 in Fig.~5
    by \citet{2006-Icarus-182-Li}
    and 
    feature b$_4$ in Fig.~7 by \citet{2008-AA-478-Carry}.}\\
  \indent \add{Feature D is one of the highest slopes measured in the 
    J-band map, although it falls on one of the main noise residuals.
    It corresponds to the dark feature with a central bright
    peak seen in the near-infrared images \citep{2008-AA-478-Carry}.
    If this round-shaped feature is associated with an impact crater,
    the observed differences in reflectivity and spectral slope may 
    be related to differences in surface age (``fresher'' material
    excavated), or regolith properties (different packing).
    Given the relative lack of constrains on the surface composition
    of Ceres, we cannot assert the causes of these variations.
    We only report here on the detection of spectral slope variations
    correlated to albedo markings.}

\section{Discussion\label{sec: discussion}}

  \indent \add{The handful of features detected here contrasts with the wealth of
    albedo markings reported by}
    \rem{The overall absence of spectral variations appears in
  contradiction with reports from }
  \citet{2006-Icarus-182-Li}
  and
  \citet{2008-AA-478-Carry}
  based on broad-band photometry in the
  visible and near-infrared
  with HST/ACS and Keck/NIRC2, respectively.
  However, both analyses used deconvolution techniques, which 
  enhance the spatial resolution and photometric contrast.
  The data presented here still suffers from aberrations
  introduced by the not fully-corrected atmospheric turbulence,
  and non-perfect optics of the instrument and telescope.
  The \add{encircled energy within a} resolution element (the
  core of the PSF, corresponding to about \add{75}\,km on the surface of
  Ceres) will thus only be a portion of the total energy.\\
  \indent The information is thus \add{spread out and averaged
  with nearest-neighbor pixels}, lowering the photometric contrast
  \citep[albedo variations are barely distinguishable on non-deconvolved
  images of Ceres used by][\add{although being perfectly
      distinguishable after deconvolution}]{2008-AA-478-Carry}. 
  This highlights the need for deconvolution algorithms for
  spectro-imaging.\\
%
%

  \indent Nevertheless, the apparent homogeneity of the surface
  composition of Ceres might be consistent with the hypothesis of a
  differentiated interior. 
  \citet{2005-JGR-110-McCord} have suggested that a thin
  ice-silicate crust would be probably unstable if it overlays a less
  dense liquid-water layer. In this case, the crust would tend to break
  up and founder probably soon after the liquid-water mantle formed in
  the first tens of Myr after formation. A new crust would then quickly
  freeze out and thus resurface Ceres by mixing and/or depositing
  minerals on the surface, erasing major albedo and morphological
  features \citep{2006-Icarus-182-Li, 2008-AA-478-Carry}.
  If this mixing was efficient, then
  the surface of Ceres may have acquired a homogeneous composition at
  early epochs of its evolution. Alternatively, this homogeneity could
  be also explained by resurfacing processes induced by (ancient or more
  recent) cryovolcanism \citep{2010-Icarus-205-Castillo-Rogez}.
  In both scenarios, the short sublimation timescale of the ice
  brought to the surface of the asteroid makes its detection unlikely.

\section{Conclusion}

  We present the first disk-resolved spectroscopic observations
  of dwarf-planet (1) Ceres in the near-infrared
  (1.1--2.4\,$\mu$m). 
  Our observations used SINFONI on the ESO Very Large Telescope and
  had an angular resolution of about 0.040\arcsec,
  corresponding to $\sim$\add{75}\,km on the surface of Ceres, and a
  spectral resolving power of \add{about} 1500.\\
  \indent \add{We did not detect any absorption band at the
  3.0\%, 1.2\%, and 1.3\% levels in J, H and K bands, respectively.
  Variations of the spectral slope are observed over the surface, with
  dark and bright albedo markings 
  \citep[few percent variations detected from visible and near-infrared
    imaging, see][]{2006-Icarus-182-Li,2008-AA-478-Carry} 
  presenting respectively a lower and higher spectral slope (couple of
  percent per micron) than the surroundings. 
  The surface of Ceres is thus remarkably homogeneous at
  our spatial and spectral resolutions.}

  \rem{All the spectral variations detected across the apparent disk
  appear to be
  linked to slope variations due to the observing geometry, and are
  uncorrelated to surface features.
  The only place where the surface may show genuine slope
  variation corresponds to the location of the 
  large dark albedo patch reported by \citet{2008-AA-478-Carry} from
  near-infrared images.
  We did not detect any absorption band at  
  3.0\%, 1.2\%, and 1.3\% level in J, H and K bands, respectively.
  The surface of Ceres is thus remarkably homogeneous at
  our spatial and spectral resolutions.}

\section*{Acknowledgments}

  This research used IMCCE's Miriade VO tool and 
  NASA's Astrophysics Data System.
  A great thanks to all the developers and
  maintainers.
  BC and PV thank the ESA Visiting Scientist Program.
  \add{
    EJ and JM are Research Associate and 
    Research Director of the FRS-FNRS, Belgium.}
  \add{We thank both anonymous referees for their valuable comments
    that helped us in improving this manuscript.}

  \bibliographystyle{icaruslike}

\end{document}